\documentclass[prb,aps,reprint,superscriptaddress,showpacs,floatfix]{revtex4-1}
\usepackage{amsmath}
\usepackage{epsfig}
\usepackage{color}
\usepackage[bookmarks=true,colorlinks=true,urlcolor=blue,linkcolor=blue,citecolor=blue]{hyperref}
\bibpunct{[}{]}{,}{n}{}{}

\begin{document}

\title{Femtosecond laser driven molecular dynamics on surfaces: Photodesorption of molecular oxygen from Ag(110)}
\author{Ivor Lon\v{c}ari\'c}
\email{ivor.loncaric@gmail.com}
\affiliation{Centro de F\'{\i}sica de Materiales CFM/MPC (CSIC-UPV/EHU), P. Manuel de Lardizabal 5, 20018 Donostia-San Sebasti\'an, Spain}
\author{M.~Alducin}
\affiliation{Centro de F\'{\i}sica de Materiales CFM/MPC (CSIC-UPV/EHU), P. Manuel de Lardizabal 5, 20018 Donostia-San Sebasti\'an, Spain}
\affiliation{Donostia International Physics Center DIPC, P. Manuel de Lardizabal 4, 20018 San Sebasti\'an, Spain}
\author{P.~Saalfrank}
\affiliation{Institut f\"{u}r Chemie, Universit\"{a}t Potsdam, Karl-Liebknecht-Strasse 24-25, D-14476 Potsdam, Germany}
\affiliation{Donostia International Physics Center DIPC, P. Manuel de Lardizabal 4, 20018 San Sebasti\'an, Spain}
\author{J. I.~Juaristi}
%
\affiliation{Departamento de F\'{\i}sica de Materiales, Facultad de Qu\'{\i}micas, Universidad del Pa\'{i}s Vasco (UPV/EHU), Apartado 1072, 20080 San Sebasti\'an, Spain}
\affiliation{Centro de F\'{\i}sica de Materiales CFM/MPC (CSIC-UPV/EHU), P. Manuel de Lardizabal 5, 20018 Donostia-San Sebasti\'an, Spain}
\affiliation{Donostia International Physics Center DIPC, P. Manuel de Lardizabal 4, 20018 San Sebasti\'an, Spain}
\begin{abstract}
We simulate the femtosecond laser induced desorption dynamics of a diatomic molecule from a metal surface by including the effect of the electron and phonon excitations created by the laser pulse. Following previous models, the laser induced surface excitation is treated through the two temperature model, while the multidimensional dynamics of the molecule is described by a classical Langevin equation, in which the friction and random forces account for the action of the heated electrons. In this work, we propose the additional use of the generalized Langevin oscillator model to also include the effect of the energy exchange between the molecule and the heated surface lattice in the desorption dynamics. The model is applied to study the laser induced desorption of O$_2$ from the Ag(110) surface, making use of a six-dimensional potential energy surface calculated within density functional theory. Our results reveal the importance of the phonon mediated process and show that, depending on the value of the electronic density in the surroundings of the molecule adsorption site, its inclusion can significantly enhance or reduce the desorption probabilities.
\end{abstract}


\onecolumngrid
\vspace{\columnsep}
Published in~Phys.~Rev.~B~93,~014301~(2016)~DOI:\href{http://dx.doi.org/10.1103/PhysRevB.93.014301}{10.1103/PhysRevB.93.014301}
\vspace{\columnsep}
\twocolumngrid

\maketitle

%
%
\section {Introduction}\label{intro}
Laser driven photochemistry has proven to be a useful tool for promoting reactions at surfaces or even as a way to open new reaction channels not accessible by thermal activation~\cite{Ho1996166, Bonn13081999,peter_rew1, peter_rew2, peter_rew3}. 
In particular, one important reaction is the photodesorption of
a molecule from a metal surface.
Generally, desorption on metals can be induced either by directly
exciting the molecule (IR photons) or it can be substrate mediated
(UV/Vis/Near IR photons). Among substrate mediated processes, one
usually distinguishes between desorption induced by electronic
transitions (DIET) and desorption induced by multiple electronic
transitions (DIMET)~\cite{misewich:90}. DIET is practically realized by
using continuous wave or nanosecond-pulse lasers with low intensity,
resulting in small desorption yields that increase linearly with laser
fluence. In DIET on metals, the adsorbate captures a hot electron and
forms a short lived excited state (negative ion resonance). After
decaying to the electronic ground state, the adsorbate may gain enough
energy and desorb. On the contrary, DIMET, which is the subject of the
present study, is realized by intense femtosecond laser pulses. Such pulses are short in comparison to typical relaxation times of adsorbate excited states and, consequently, they can produce multiple excitations of an adsorbate that lead to desorption. DIMET results in relatively large desorption yields that increase superlinearly with laser fluence~\cite{Ho1996166}.

Different methods have been used to model DIMET~\cite{peter_rew1,peter_rew2,peter_rew3}.~Several of them are the so-called excitation-deexcitation models, in which the system jumps between two or more electronic states (see review~\cite{peter_rew1} for a complete list). However, these methods due to their complexity have only been applied to a reduced number of degrees of freedom. 
In this work, we use an alternative model~\cite{tullyCOcu100,CO/Cu100,
luntz:06, peter_pccp} that permits treating all the molecular degrees of
freedom. Instead of treating excited states explicitly, in this model
the nuclear motion is classical in the ground state potential and all
the electronic degrees of freedom are included via friction and
associated fluctuation forces. 
The friction force accounts for the dissipation of the adsorbate energy
on the surface by creation of low energy electron-hole pairs, while
fluctuation forces represent the inelastic scattering of hot electrons on the adsorbate nuclei. The magnitude of the fluctuation force is obtained in terms of the temperature of the laser-induced hot electrons. This electronic temperature can be estimated from the properties of the laser pulse and the metal substrate. The first important ingredient of this model is an accurate ground state potential, which 
can be modelled with a range of methods with increasing accuracy and
theoretical, as well as, computational complexity, starting with simple
two body potentials up to accurate quantum chemistry methods. Early works that used the molecular dynamics with electronic friction model to simulate the laser-induced desorption were based on empirical potentials~\cite{tullyCOcu100,CO/Cu100}. Nowadays, one can obtain better accuracy and predictability by state of the art non-empirical theoretical methods. Particularly, a good balance between accuracy and computational complexity is achieved by density functional theory (DFT). This method, already at its semi-local level, is able to capture reasonably well both metallic delocalized states and molecular localized states and their interaction.

Ab initio molecular dynamics, in which DFT is used at each integration step to
calculate the forces, keeps both the DFT accuracy and the full
dimensionality of the problem. However, it is still computationally too
demanding to treat low probability processes or even to run long time (more than
few ps) dynamics. In this work we are interested in
phenomena that typically demand both large statistics and long time dynamics.
Therefore, it is more advantageous to follow an alternative scheme that
consists in constructing the adiabatic potential energy surface
(PES) from a large set of DFT energies.
Several interpolation techniques have been developed to obtain PESs
that preserve the DFT accuracy~\cite{CRP, shepardint, Lorenz2004210}. The main
prerequisite is to reduce the dimensionality of the problem to the
molecular degrees of freedom, to decrease the computational demand, thus
keeping the surface frozen. This means that in the case of diatomic
molecules, the interaction of the molecule and the surface is described
by a six dimensional (6D) PES. This approach has been successfully
applied to study the dynamics of different molecules on different metal
surfaces~\cite{Busnengo2002515, PhysRevLett.96.096102, PhysRevLett.97.056102,
Diaz06112009,PRLitziar}.

Recently, the laser-induced associative desorption of H$_2$ on Ru(0001)
has been successfully modelled by using such DFT-based 6D PES~\cite{peter_pccp}. In that work, the metal surface is kept
frozen and the laser excitation is only modelled by random scattering of
hot electrons with the nuclei of the molecule. Here, we extend this
model by allowing for lattice movement that enables us to incorporate
laser-induced phonon excitations. The study of the effect of phonons in
photodesorption, compared to that of electronic excitations, is one of
the main objectives of this work.

As an example, we will employ this methodology to study the
laser-induced desorption of O$_2$ on Ag(110). Due to the importance of
oxygen adsorption on silver surfaces in the process of ethylene
epoxidation, this system has been a subject of numerous studies~\cite{rocca,kleyn,bartolucci,review,ho1,gravilprl}. Using the \emph{corrugation reducing
procedure} (CRP)~\cite{CRP}, we have recently constructed the first
DFT-based 6D PES for O$_2$ on Ag(110) and used it to study
the dissociative adsorption process~\cite{ivor1}. The O$_2$ molecule can
adsorb on Ag(110) on several adsorption sites that are characterized by
different adsorption energies and electronic densities and, as such, it is an interesting model system. It gives us the possibility to investigate the importance of including the phonon excitations in the model for desorption from adsorption wells of different characteristics.

Photochemistry of O$_2$ on Ag(110) after substrate mediated photoexcitation under DIET conditions has been studied experimentally in Refs.~\cite{zhu:1991,ho:1994,zhu:1996}. Photodesorption, photodissociation, and also CO$_2$ formation were observed there. To our knowledge, no experimental studies under DIMET (femtosecond laser) conditions have been carried out so far. As such, our investigation has a strong predictive character.

The paper is organized as follows. The theoretical model and its implementation are described in Sec.~\ref{traj}. Application of this model to the desorption of O$_2$ from Ag(110) is examined in Sec.~\ref{results}. The main conclusions of the paper are summarized in Sec.~\ref{summary}.
%
%
\section {Theoretical model}\label{traj}
\subsection{Langevin dynamics for laser-driven nuclear dynamics at surfaces}
The response of a metal surface to the excitation generated by an
ultra short laser pulse can be described by the so called \emph{two
temperature model} (2TM)~\cite{rusi_2tm}. In this model, the
equilibration between the electron and lattice heat baths with
temperatures $T_{el}$ and $T_{ph}$, respectively, is described by the
following coupled diffusion equations, 
\begin{equation}\label{eq:2tm1}
C_{el}\frac{\partial T_{el}}{\partial t}=\frac{\partial}{\partial z}\kappa\frac{\partial T_{el}}{\partial z}-g(T_{el}-T_{ph})+S(z,t),
\end{equation}
\begin{equation}\label{eq:2tm2}
C_{ph}\frac{\partial T_{ph}}{\partial t}=g(T_{el}-T_{ph}),
\end{equation}
where $C_{el}$ is the the electron heat capacity, $C_{ph}$ is the phonon
heat capacity, $\kappa$ is the electron thermal conductivity, $g$ is the
electron-phonon coupling constant, and $S(z,t)$ is the laser source term.
In the regime of intense laser pulses that are studied here, metal
electrons are rapidly heated to several thousands K due to the low
electron heat capacity $C_{el}$ of metals. The formed hot electrons can either diffuse to the bulk [first term in the right hand side of Eq.~\eqref{eq:2tm1}] or transfer heat to the lattice phonons [term $g(T_{el}-T_{ph})$ in Eqs.~\eqref{eq:2tm1}~and~\eqref{eq:2tm2}].
The heat source term $S(z,t)$ is calculated for a metal film of thickness $d$ by
\begin{equation}\label{eq:2tm3}
S(z,t)=\frac{I(t)e^{-\alpha z}}{1-e^{-\alpha d}},
\end{equation}
where $I(t)$ is the adsorbed fraction of a laser pulse intensity and $\alpha^{-1}$ is the optical penetration depth. The latter is calculated from the laser wavelength $\lambda$ and the imaginary part of the refractive index of the surface $k$ as $\alpha^{-1}=\lambda/(4\pi k)$.

Following Ref.~\cite{tullyCOcu100}, the laser-induced dynamics of the adsorbed molecule is modelled using a Langevin equation for each atom $i$ in the molecule,
\begin{eqnarray}\label{eq:langevin}
m_i\frac{d^2\mathbf{r}_i}{dt^2}=&-&\nabla_i V(\mathbf{r}_i-\mathbf{r}_s,\mathbf{r}_j-\mathbf{r}_s)-\eta_{el,i}(\mathbf{r}_i-\mathbf{r}_s)\frac{d\mathbf{r}_i}{dt} \nonumber \\
&+&\mathbf{R}^{el}_i(T_{el},\eta_{el,i}(\mathbf{r}_i-\mathbf{r}_s)), \quad i\neq j,
\end{eqnarray}
where $m_i$, $\mathbf{r}_i$, and $\eta_{el,i}$ are their corresponding mass, position vector, and electronic friction coefficient (for diatomic molecules as studied here $i,j=1,2$). The first term on the right hand side of the equation is
the adiabatic force exerted on each atom of the molecule that originates from the interaction between the molecule and the surface. Here, this force is calculated as the gradient over the atomic coordinates of the 6D DFT PES. Note that the positions of the atoms are set relative to the position of the surface topmost layer $\mathbf{r}_s$. Surface movement $\mathbf{r}_s(t)$ and the subsequent energy transfer between the molecule and the lattice atoms are calculated within the generalized Langevin oscillator (GLO) model~\cite{glo3,glo1,glo2} as explained below in Sec.~\ref{GLO}.
Non-adiabatic effects due to the coupling of the atoms in the molecule with the surface electronic excitations enter Eq.~\eqref{eq:langevin} through the friction force $-\eta_{el,i}\frac{d\mathbf{r}_i}{dt}$ and the random fluctuating force $\mathbf{R}^{el}_i$, as described in  Sec.~\ref{LDFA}.

\subsection{Phonon excitations: The generalized Langevin oscillator model}\label{GLO}
As discussed in Sec.~\ref{intro}, the use of a 6D PES in the dynamics
equations limits the possibility of including the dynamical energy exchange between the molecule and the
surface lattice, i.e., of allowing phonon
excitations/deexcitations. One successful model that is able to keep the
accuracy of a DFT based PES and at the same time provides a reasonable description of the surface movement is the generalized Langevin oscillator (GLO) model~\cite{glo3,glo1,glo_prl,glo2}.
In the GLO model, surface motion is described in terms of a three
dimensional (3D) harmonic oscillator of mass $m_s$ with position vector
$\mathbf{r}_s$ and associated diagonal $3\times 3$ frequency matrix
$\Omega^2$. Energy dissipation and thermal fluctuations are modelled with the
help of a ghost 3D oscillator with position vector $\mathbf{r}_g$. 
The mass and the associated frequency matrix for the ghost oscillator are also $m_s$ and $\Omega^2$. 
The equations of motion for the surface and ghost oscillators, which are
coupled by the coupling matrix  $\Lambda_{gs}$, are the following,
\begin{equation}\label{eq:glors}
m_s\frac{d^2\mathbf{r}_s}{dt^2}=-\nabla_s V(\mathbf{r}_i-\mathbf{r}_s,\mathbf{r}_j-\mathbf{r}_s)-m_s\Omega^2\mathbf{r}_s+m_s\Lambda_{gs}\mathbf{r}_g,
\end{equation}
\begin{equation}\label{eq:glorg}
m_s\frac{d^2\mathbf{r}_g}{dt^2}=-m_s\Omega^2\mathbf{r}_g+m_s\Lambda_{gs}\mathbf{r}_s-\eta_{ph} \frac{d\mathbf{r}_g}{dt}+\mathbf{R}^{ph}(T_{ph}).
\end{equation}
The friction force $-\eta_{ph} \frac{d\mathbf{r}_g}{dt}$ models energy
dissipation from the interacting surface atoms to the bulk thermal bath.
Following Ref.~\cite{glo1}, the friction coefficient $\eta_{ph}$ is
calculated from the Debye frequency $\omega_D$ as $\eta_{ph}=m_s \pi
\omega_D /6$. The random fluctuation force $\mathbf{R}^{ph}$, which
models the heating of the surface atoms due to the thermal motion of the
bulk atoms, is a Gaussian white noise with variance
\begin{equation}\label{eq:gloW}
Var(\mathbf{R}^{ph}(T_{ph}))=\sqrt{\frac{2 k_B T_{ph}(t) \eta_{ph}}{ \Delta t}},
\end{equation}
where $\Delta t$ is the time integration step, $k_B$ is the Boltzmann
constant, and $T_{ph}$ is the time-dependent phonon (surface)
temperature that is calculated in the 2TM model. The friction and
random fluctuation forces are linked by the fluctuation-dissipation theorem to
ensure that the surface atoms are coupled to a thermal bath of $T_{ph}$.
Oscillator frequencies $(\Omega^2)_{ii}=2 \omega_i^2$ and coupling matrix elements
$(\Lambda_{gs})_{ii}=\omega_i^2$ are obtained from the surface phonon frequencies $\omega_i\,(i=x,y,z)$ at the edges of the surface Brillouin zone, as implemented in Refs.~\cite{glo_prl,glo2}.

\subsection{Electronic non-adiabatic effects. Local density friction
approximation}\label{LDFA}
The Born-Oppenheimer (adiabatic) approximation, in which the electrons react instantaneously to the nuclear motion, is a cornerstone in gas-surface dynamics. Nevertheless, the existence of a nonadiabatic energy dissipation upon adsorption of gas species (atomic or molecular) on metal surfaces through electron-hole pair excitations is well established~\cite{Gergen21122001, PhysRevB.75.073410}. Several methods have been used to model this dissipation mechanism~\cite{luntz:05, Lindenblatt20063624, PhysRevLett.100.116102, Shenvi06112009, kratzer}. Among them, a method that has proven to be both accurate and suitable to perform multidimensional molecular dynamics is the \emph{local density friction approximation} (LDFA)~\cite{PhysRevLett.100.116102}. In this model, electronic non-adiabatic dissipative effects are introduced in the dynamics via a friction force proportional to the velocity of the atom, as in Eq.~\eqref{eq:langevin}. The friction coefficient $\eta_{el}$ is obtained in terms of the scattering of electrons by an atom inside a homogeneous free electron gas (FEG) as
\begin{equation}
\eta_{el}=\frac{4\pi n}{k_F}\sum_{l=0}^{\infty}(l+1)\sin^2[\delta_l(k_F)-\delta_{l+1}(k_F)].
\label{LDFAequa}
\end{equation}
In this equation, $n$ is the FEG density and $k_F$ is the Fermi momentum. The $\delta_l(k_F)$ are the scattering phase shifts evaluated at the Fermi level corresponding to the potential induced by the atom in the FEG, which is calculated within DFT.
The friction coefficient of Eq.~\eqref{LDFAequa} has successfully been used to calculate the stopping power of atoms and ions in metal solids and surfaces~\cite{ECHENIQUE1981779, PhysRevA.33.897, PhysRevB.67.245401, PhysRevLett821048}.
Within the LDFA, the electronic density entering Eq.~\eqref{LDFAequa} is
chosen at each point of the trajectory as that of the bare surface at
the position of the atomic nuclei $n(\mathbf{r}_i-\mathbf{r}_s)$. The latter can be easily obtained from a DFT calculation as described in Sec.~\ref{impl} below. 
The LDFA has been applied to study the effect of electronic excitations
in the dynamics of atoms and molecules on metal surfaces~\cite{PhysRevLett.100.116102, PhysRevLett108096101, itziar_ldfa, PhysRevB.81.125408, PhysRevLett112103203, peter_aimdef, jpcc_er}.

The scattering of heated electrons with the adsorbate results also in a
fluctuating force $\mathbf{R}^{el}_i$ [see Eq.~\eqref{eq:langevin}] that is connected through the fluctuation-dissipation theorem to the electronic friction force 
via the electronic temperature $T_{el}$~\cite{tullyCOcu100}. Here, $\mathbf{R}^{el}_i$ is modelled by a Gaussian white noise with variance
\begin{equation}\label{eq:fluc}
Var(\mathbf{R}^{el}_i(T_{el}))=\sqrt{\frac{2 k_B T_{el}(t) \eta_{el,i}}{\Delta t}}.
\end{equation}
Note that $\mathbf{R}^{el}_i$ is usually negligible for the typical thermal surface temperatures used in gas/surface experiments and it can safely be neglected. However, this term gives a large contribution in case of the high $T_{el}$ that are obtained in the wake of the laser excitation, particularly, for adsorbates embedded in high electron density regions of the surface.

\subsection{Implementation of the method}\label{impl}
We start by solving the 2TM differential Eqs.~\eqref{eq:2tm1}-\eqref{eq:2tm3} to obtain $T_{el}$ and $T_{ph}$ as a function of time for the specific surface and laser pulse properties of interest.
The calculated time dependent electronic and phonon temperatures are
saved on a grid (in practice in steps of $0.05$ ps) and used as inputs
in the molecular dynamics calculations [Eqs.~\eqref{eq:langevin}-\eqref{eq:fluc}]. Another required input that is
needed to obtain $\eta_{el}$ is the electronic density of the bare
surface $n(\mathbf{r})$. Here, $n(\mathbf{r})$ is calculated with DFT and saved on a real space grid.

We perform classical dynamics calculations that neglect the zero point
energy of the adsorbate. Each trajectory starts with the molecule
resting in one of the adsorption wells. The initial position of the
surface $\mathbf{r}_s$ (and the corresponding momenta) are sampled by a
conventional Monte Carlo procedure, such that they correspond to the
initial surface temperature. The dynamics equations
\eqref{eq:langevin}-\eqref{eq:glorg} are integrated with a Beeman
algorithm~\cite{BEEMAN1976130} as implemented in Refs.~\cite{beeman2,
glo2}. At each integration step, the corresponding $T_{el}$ and $T_{ph}$ are obtained by a cubic spline interpolation. The electronic density 
at the position of each atom in the molecule $n(\mathbf{r}_i)$ is obtained with a 3D cubic spline interpolation of the DFT calculated bare surface density.

Using the same implementation that solves Eqs.~\eqref{eq:langevin}-\eqref{eq:glorg}, one can also perform dynamics simulations that only include the electronic or the phonon contribution by setting, respectively, $\mathbf{r}_s=0$ or $\eta_{el,i}=0$ in Eq.~\eqref{eq:langevin}. In the following, the three types of calculations will be denoted as, LDFA+GLO, when including both the electronic and phonon contributions, LDFA, when including only the electronic channel, and GLO, when only phonons are included.
%
%
\section {Laser induced desorption of O$_2$ from A\lowercase{g}(110)}\label{results}
\subsection{System properties: Results from DFT calculations}
The ground state properties of O$_2$ on Ag(110) are described by a recently constructed 6D PES~\cite{ivor1} that is obtained from the CRP interpolation of $\sim 25000$ spin-polarised DFT energies. The latter were calculated with the Perdew-Burke-Ernzerhof (PBE) exchange correlation functional~\cite{PBE} as implemented in the VASP code~\cite{VASP1, VASP2} and with a plane-wave basis set energy cut-off of $400$~eV. The surface was modelled by a supercell consisting of a ($2\times3$) surface unit cell, a five-layer thick slab, and 14 layers of vacuum. Additional details are given in Ref.~\cite{ivor1}.

\begin{figure}[ht]
\includegraphics[width=0.9\columnwidth]{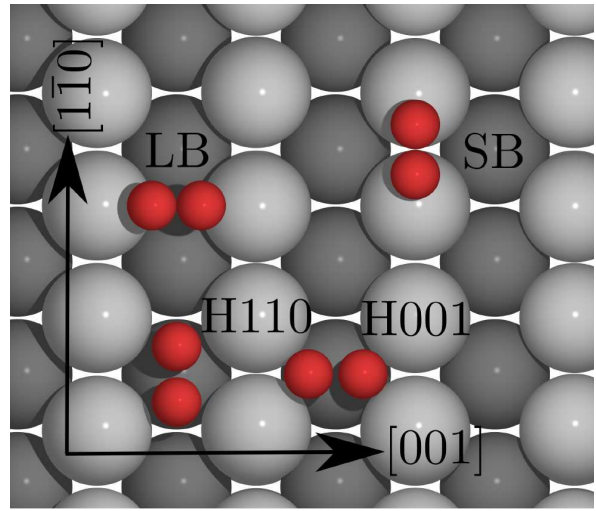} 
\caption{(Color online) Sketch of the position of the molecule in the four adsorption wells as predicted by DFT: long bridge (denoted LB), short bridge (denoted SB), hollow along the [1\=10] direction (denoted H110), and hollow along the [001] direction (denoted H001). First layer surface atoms are shown in light grey, while second layer atoms are shown in dark grey.\label{pos}}
\end{figure}

The oxygen molecule on Ag(110), as predicted by DFT with the PBE
exchange-correlation functional, features the four adsorption wells
sketched in Fig.~\ref{pos}: 
In the SB well, the molecular center of mass (CM) is at the short bridge site with the molecular axis
oriented along the [1\=10] direction. In the LB well, the molecular CM is at the
long bridge site with the molecular axis oriented along the [001]
direction. In the H001 well, the molecular CM is at the hollow site with the
molecular axis oriented along the [001] direction. In the H110 well, the molecular CM
is at the hollow site with the molecular axis oriented along the [1\=10]
direction. In all these adsorption sites the molecule is parallel to the
surface.

\begin{table}[ht]
\caption{Properties of the adsorption states of O$_2$ on Ag(110): Adsorption energy $E_a$,
 O$_2$-surface distance $Z$, interatomic O-O distance $r$, and electronic
 density in which oxygen atoms are embedded (expressed in terms of the
 mean free electron radius $r_s$ given in atomic units, a.u.). \label{tableADS}}
\begin{ruledtabular}
\begin{tabular}{c c c c c c}
Site & $E_a$ (eV) & $Z$ (\AA) & $r$ (\AA) & $r_s$
(a.u.) & \\ \hline
LB & $-0.24$ & 1.98 & 1.29 & 3.82 \\
SB & $-0.33$ & 2.20 & 1.31 & 3.57 \\
H001 & $-0.24$ & 1.29 & 1.42 & 2.62 \\
H110 & $-0.21$ & 1.09 & 1.45 & 2.57\\ 
\end{tabular}
\end{ruledtabular}
\end{table}

Table~\ref{tableADS} summaries the main features of each adsorption
position, namely, the adsorption energy $E_a$, the distance $Z$ from the
surface of the molecular CM, the O$_2$ internuclear distance $r$, and
the value of the bare surface electron density at the position of each O
atom, which is given in terms of the mean free electron radius $r_s$.
The H110 well is the closest to the surface and its adsorption energy is
$E_a=-0.21$~eV. The H001 well is somewhat further from the surface, but
with a larger adsorption energy of $-0.24$~eV.
It is important to remark that all these values correspond to the
results obtained with the frozen surface 6D PES that is used in our
simulations of Sec.~\ref{res_and_diss}. However, we have checked that if
the surface is allowed to relax, the H001 and H110 adsorption energies
increase (in absolute value) to a similar value of around $-0.36$~eV. Note that the DFT-PBE
description of the H001 and
H110 adsorption wells seems to be in good agreement
with experimental observations~\cite{ho1,outka,guest,review}. 

Adsorption wells at the bridge sites (LB and SB) are further away from
the surface compared to the hollow wells. 
Considering the $Z$ values of the bridge wells, one could be tempted to
assign them to the measured physisorption
wells~\cite{tangphysisorbed,bartolucci,review}. However,
Table~\ref{tableADS} shows that DFT-PBE predicts too large adsorption
energies in these wells to be considered as physisorbed states~\cite{gravilprl}.
In spite of this, our study is meaningful since it uses a state of the art PES for O$_2$ on Ag(110).
Additionally, the use of a PES for a system that presents several adsorption sites with different characteristics is advantageous for a theoretical study over systems with just one adsorption site. Having one system with several wells gives us the opportunity to more clearly study the dependence of the results on the properties of the wells, such as the adsorption energy, the distance from the surface, and thus, the electron density in which the molecule is embedded. In this respect, our results can be predictive for systems in which adsorption wells with similar characteristics exist.

\subsection{Computational details}

Our simulations are performed for laser pulses of Gaussian shape with $800$~nm wavelength, $130$~fs of full width at half maximum (FWHM), and absorbed fluences in the range $F=50-200$~J/m$^2$. Laser pulses with these properties were used in desorption experiments performed on other systems~\cite{Bonn13081999,PhysRevLett.91.226102}. The laser-induced $T_{el}$ and $T_{ph}$ are calculated using the following  material constants for Ag:
$C_{el}=63.3$~J/m$^3$K, $\kappa=429$~W/mK, $\gamma=63.3$~J/m$^3$K, and $k=5.29$~\cite{PhysRevB.77.075133,majchrzak2010two,palik1985handbook}. The phonon heat capacity $C_{ph}$ is calculated in the Debye model, with Debye temperature $T_D(\mathrm{Ag})=225$~K. The metal slab thickness $d$ in Eq.~\eqref{eq:2tm3} is set to 0.5~$\mu$m. We have checked that with this $d$-value the calculated $T_{el}$ and $T_{ph}$ are well converged.

The GLO equations for the Ag(110) surface are solved using $\omega_x=\omega_y=3.7\times 10^{-4}$~a.u. (atomic units) and $\omega_z=2.9\times 10^{-4}$~a.u. for the surface oscillator frequencies~\cite{Bracco1997325, Narasimhan2002331}, and $\eta_{ph}=74.4$~a.u. for the friction coefficient of the ghost oscillators.

The electronic friction coefficient entering Eqs.~\eqref{eq:langevin}
and \eqref{eq:fluc} as a function of the embedding density is given by
\begin{equation}
\eta_{el}(r_s)=1.365\, r_s^{-1.828} e^{-0.082 r_s}+50.342\, r_s^{0.491}e^{-2.704r_s},
\end{equation}
where both $r_s$ and $\eta_{el}$ are in a.u. 
This function fits the friction coefficients of an
oxygen atom calculated for embedding FEG densities varying in the range $r_s=1-6$~a.u. This range covers all the electronic density values that are
relevant in our dynamics.

In all the calculations presented below, the initial surface temperature is set to $T_{ini}=100$~K. To enable the thermalization of the molecule prior to the laser excitation, the laser pulse is turned on after 1.5 ps, thus keeping the initial temperature constant during this time interval. However, we have checked that the results of the dynamics do not depend on this thermalization time and that the laser pulse could be turned on at the beginning of the dynamics calculation without altering the final results. We have also checked that the largest integration step that can be used keeping the results of the dynamics stable is 1 fs. In all cases the integration time is $50$~ps and the instant $t=0$ corresponds to the start of the trajectory calculation.

As an outcome of our dynamics we consider that a molecule has been
desorbed when its center of mass arrives at $6$~\AA~from the surface and its velocity direction points away from the surface. We also distinguish another possible
exit channel, dissociation, if the interatomic distance $r$ is larger than
$2.5$~\AA~ with positive radial velocity. 

\subsection{Results and discussion} \label{res_and_diss}
\begin{figure}[ht]
\includegraphics[width=\columnwidth]{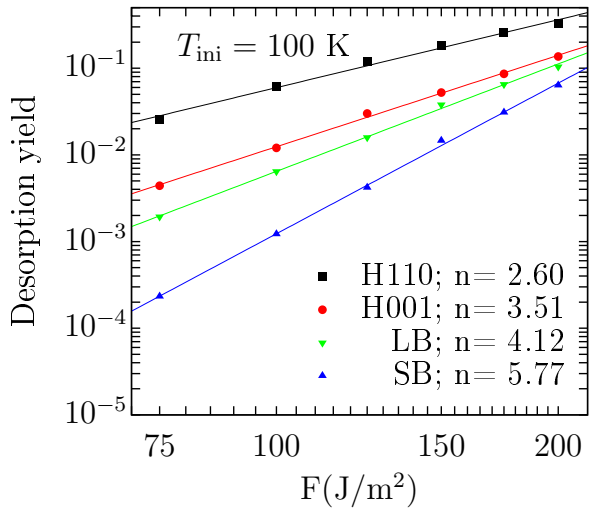} 
\caption{(Color online) Desorption yields $Y$ from the four adsorption wells (shown with different symbols and colors) as a function of the laser fluence $F$. For every well the coefficient $n$ is calculated by fitting the data to the equation $Y=aF^n$.\label{best}}
\end{figure}

Calculated desorption yields $Y$ as a function of the laser fluence $F$ are shown in Fig.~\ref{best} for the four different adsorption wells. These values have been obtained from the number of desorption events out of 30\,000 trajectory calculations performed for each laser fluence and initial adsorption position. 
Characteristic super-linear desorption yields, which follow a power law
$Y=aF^n$ with $n>1$, are observed for the four wells. The values of the
exponent $n$ are in the range $2.6-5.8$. These values are similar to
those obtained for other systems~\cite{peter_rew1}, such as CO/Cu(100)
with $n=5-8$~\cite{CO/Cu100,PhysRevLett.77.4576}, CO/Pd(111) with $n=7-9$~\cite{CO/Pd111}, NO/Pt(111) with $n=6$~\cite{Ho1996166,
NO/Pt111}, O$_2$/Pt(111) with $n=6$~\cite{PhysRevLett.71.2094,O2/Pt111}, O$_2$/Pd(111) with $n=6-9$~\cite{O2/Pd111}, and associative
desorption of H$_2$/Ru(0001) with $n=3$~\cite{PhysRevLett.91.226102,peter_pccp}. Independent
of the considered laser fluence, the highest desorption yields are
obtained for H110, followed by H001, LB, and SB. The exponent $n$ of the power law is also different for each well, its value decreasing from $n=5.8$ for desorption from the SB well down to $n=2.6$ for desorption from the H110 well. Both results can be mostly related to the differences in the adsorption energies of the different wells (see $E_a$ in Table~\ref{tableADS}). The highest desorption yield and lowest exponent correspond to the well with the lowest adsorption energy and vice versa.
However, the adsorption energy itself is not the only property ruling
the desorption process. The LB and H001 wells have the same adsorption
energy ($-0.24$~eV), but the yields 
are consistently larger for desorption from the H001 well than from the LB well. As shown below, this effect is related to the different mechanisms that rule desorption from the hollow and the bridge sites. 

\begin{figure}[ht]
\includegraphics[width=\columnwidth]{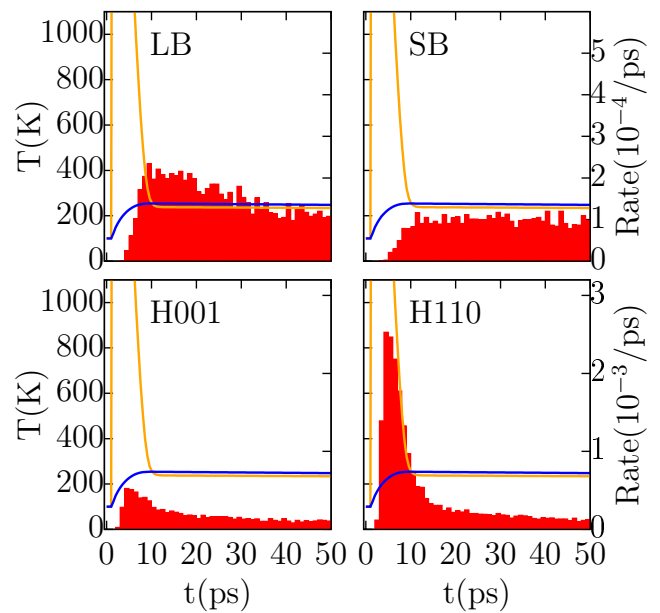} 
\caption{(Color online) Desorption rates as a function of time for $F=200$~J/m$^2$ from the four adsorption sites (right ordinate). Note the different scales used for the bridge sites (upper panels) and the hollow sites (lower panels). Electron (orange line) and phonon (blue line) temperatures calculated from the 2TM are also shown (left ordinate). The electronic temperature peaks at values $T_{el} > 6000$~K (see Fig.~\ref{rate2}). The histograms are obtained by counting desorption events in intervals of 1 ps. \label{rate1}}
\end{figure}

Figure~\ref{rate1} shows the time dependence of $T_{el}$ and $T_{ph}$ as
obtained from the 2TM for $F=200$~J/m$^2$ in comparison with the time
evolution of the desorption rate from each of the adsorption wells.
There are remarkable differences between the bridge wells (LB and SB)
and the hollow wells (H001 and H110) observed not only in the magnitude of the desorption rates, but also in their time evolution. 
The desorption rates for the hollow wells seem to follow the time evolution of $T_{el}$, but with a delay of around $3.5$~ps. 
In contrast, the desorption rates from the bridge sites do not seem to be very much affected by the high increase of $T_{el}$ at short times. In these cases, the highest values of the desorption rates occur at longer times, once $T_{el}$ and $T_{ph}$ are equilibrated.
It is worth to mention that the desorption rate from SB seems to follow the time evolution of $T_{ph}$, but also with a certain delay. 
On the one hand, these observations suggest that desorption from the hollow sites is mainly an electron mediated effect, where the energy transfer from the electrons excited by the laser pulse to the adsorbed molecule plays a dominant role. On the other hand, these results also suggest that the heating of electrons is not that important for desorption from the bridge sites and that the laser mediated phonon excitation is the relevant mechanism in this case. 
In order to confirm these ideas and gain further insight in the relative importance of the electron and phonon mediated mechanisms, we have performed the two additional types of calculations described in Sec.~\ref{impl} above, in which only the effect of either the heated electrons (LDFA) or heated phonons (GLO) is included in the desorption dynamics. 

\begin{table*}[ht]
\caption{Desorption yields from the four sites and for two different laser fluences calculated with the full model (Y$_\mathrm{LDFA+GLO}$), the model in which the surface is frozen (Y$_\mathrm{LDFA}$), and the model in which electronic excitations are neglected (Y$_\mathrm{GLO}$). \label{tableY}}
\begin{ruledtabular}
\begin{tabular}{c c c c c c c c c c}
 &  \multicolumn{4}{c}{$F=100$~J/m$^2$}  & &  \multicolumn{4}{c}{$F=200$~J/m$^2$}  \\
model  & LB  & SB & H001 & H110 &  &  LB  & SB & H001 & H110  \\ \hline
Y$_\mathrm{LDFA+GLO}$ & 0.006 & 0.001 & 0.012 & 0.062 &  & 0.104  & 0.064 & 0.136 & 0.325 \\
Y$_\mathrm{LDFA}$ & 3$\times10^{-4}$ & 6$\times10^{-5}$ & 0.088 & 0.311  & &  0.023  & 0.011 & 0.369 & 0.740 \\
Y$_\mathrm{GLO}$ & 0.007 & 7$\times10^{-4}$ & 0.007 & 0.013 & & 0.108 & 0.051 & 0.112 & 0.165 \\
\end{tabular}
\end{ruledtabular}
\end{table*}

\begin{figure}[ht]
\includegraphics[width=\columnwidth]{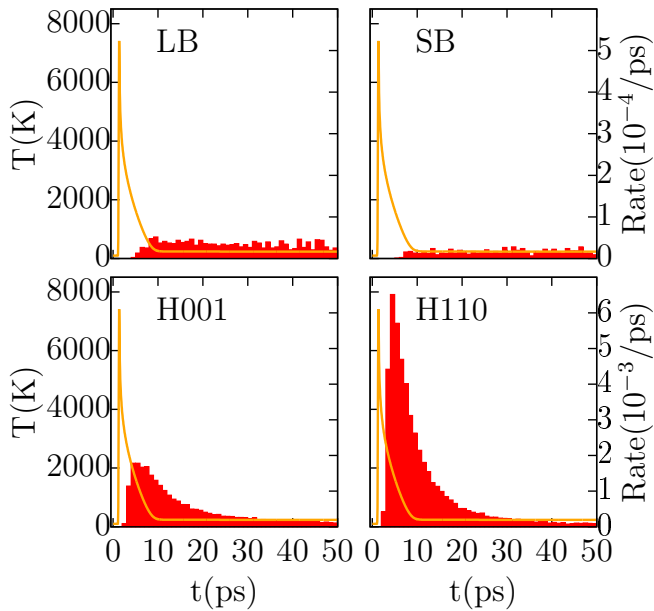} 
\caption{(Color online) Desorption rates as a function of time for $F=200$ J/m$^2$ from the four adsorption sites  calculated with the surface frozen (LDFA model) (right ordinate). Laser excitation of the surface is modelled only by the electronic temperature (orange line) given by the 2TM (left ordinate). Note the different scales used for the bridge sites (upper panels) and the hollow sites (lower panels).\label{rate2}}
\end{figure}

\begin{figure}[ht]
\includegraphics[width=\columnwidth]{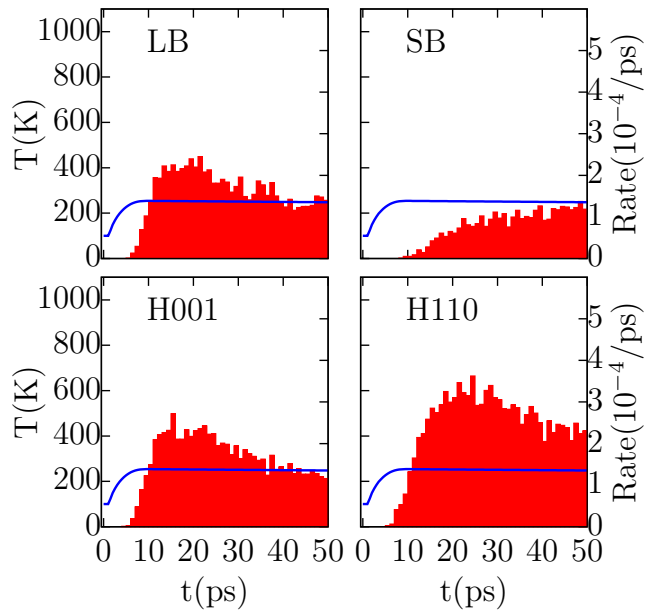} 
\caption{(Color online) Desorption rates as a function of time for $F=200$ J/m$^2$ from the four adsorption sites calculated neglecting electronic excitations (GLO model) (right ordinate). Laser excitation of the surface is modelled only by phonon temperature (blue line) as given by the 2TM (left ordinate). \label{rate3}}
\end{figure}

The desorption yield obtained from the four adsorption wells for two
different laser fluences, using the three different models (LDFA, GLO, and
LDFA+GLO), are given in Table~\ref{tableY}. Additionally, the desorption
rates for laser fluence $F=200$~J/m$^2$ calculated with the LDFA and GLO
models are shown in Figs.~\ref{rate2} and~\ref{rate3}, respectively. 
The new LDFA and GLO results confirm the ideas inferred above. In the case of
desorption from the hollow sites, the LDFA yields and rates are significantly
larger than the GLO ones, while the opposite behavior is observed for desorption
from the bridge wells although the differences between GLO and LDFA are
smaller in these cases. Focusing on the LDFA calculations, it is clear that
the desorption yields (Table~\ref{tableY}) and rates for the bridge
wells (Fig.~\ref{rate2}) are reduced to
marginal levels as compared to the ones obtained for the hollow sites. However,
Fig.~\ref{rate3} and the GLO values in Table~\ref{tableY} show that the phonon-mediated contribution
to desorption is rather similar among the four wells. In fact, the
small differences we observe 
seem to be correlated with the differences in adsorption energy. Thus, the
lowest yield corresponds to the SB site, the one with the largest $E_a$,
and the largest yield to the H110, the one with the lowest $E_a$. 
The intermediate cases represented by the LB and H001 sites, which have
the same $E_a$, show very similar desorption yields. The
absence of a similar one to one correspondence between $E_a$ and the LDFA
yields points to the electronic-mediated mechanism as the one
responsible for removing that correlation in the LDFA+GLO yields, since
in both cases the largest to lowest values for desorption follow the
order H110, H001, LB, and SB. Yet, it remains to be understood what
property (together with $E_a$) rules the efficiency of the electronic
mechanism.

\begin{figure}[ht]
\includegraphics[width=\columnwidth]{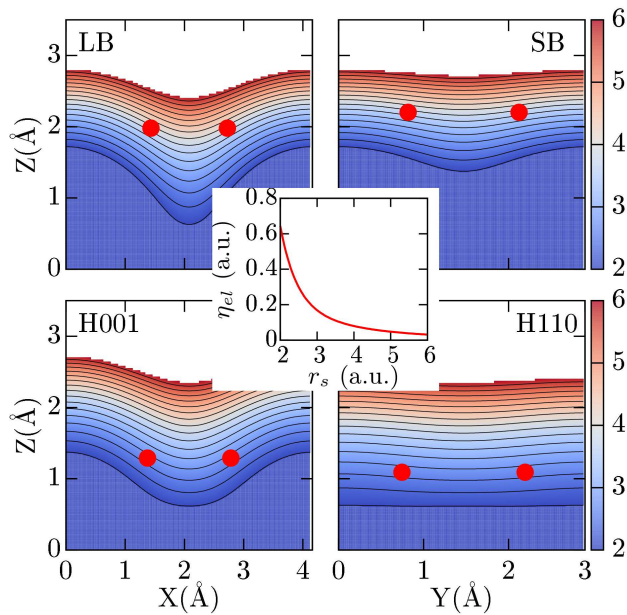} 
\caption{(Color online) Contour plot of electronic densities, expressed in terms of the mean free electron radius $r_s$, for the configurations of the four adsorption sites. Contour lines are separated by $0.25$~a.u., in the range from $2$~a.u. to $6$~a.u. as shown by the color map. The positions of the oxygen atoms in the adsorption sites are shown with red dots. The inset shows the friction coefficient as a function of electronic density given in terms of $r_s$. \label{den}}
\end{figure}

As explained in Sec.~\ref{LDFA}, the electronic contribution to desorption is determined
within the LDFA description by the value of the bare surface electron density
at the position of each adsorbate (in our case the O atoms). The density profile
along the plane normal to the surface that contains the molecule is shown in
Fig.~\ref{den} for each of the adsorption configurations, together with the corresponding
O atom positions. The inset shows the friction coefficient of one O atom as a function of
the electronic density. Clearly, the embedding electron density is higher when
the molecule is adsorbed on the hollow than on the bridge wells (see also Table~\ref{tableADS}). 
This nicely fits with the results we have obtained. When $T_{el}$ is high,
the fluctuation forces acting on O$_2$ are correspondingly larger if adsorbed on
the hollow wells than if adsorbed on the bridge wells. Therefore, despite the
similar adsorption energies of H001 and LB, desorption is more efficient from
the former because of the larger embedding density.

Further insight regarding the competition between the electron- and
phonon-mediated mechanisms can be gained by comparing the LDFA and GLO
results to those obtained with the LDFA+GLO simulations. Thus, the LDFA
model predicts larger yields for the electron-dominated desorption cases
(H110, H001) than the LDFA+GLO. The reason is that the 
adsorbed O$_2$, being efficiently heated during the initial time
interval in which $T_{el}$ is high, reaches temperatures larger than
$T_{ph}$ at least during this period. Therefore, when surface motion is
also included in the dynamics, the surface takes energy from the
electronically heated molecule and the desorption probability is reduced
in respect to the ideal case in which no surface motion is allowed.
In the case of the GLO simulations, the yields are slightly larger than the LDFA+GLO ones for the LB site, suggesting energy uptake by the electronic system, while for the SB well the LDFA+GLO yields almost coincide with the sum of the GLO and LDFA values. 

\begin{figure}[ht]
\includegraphics[width=\columnwidth]{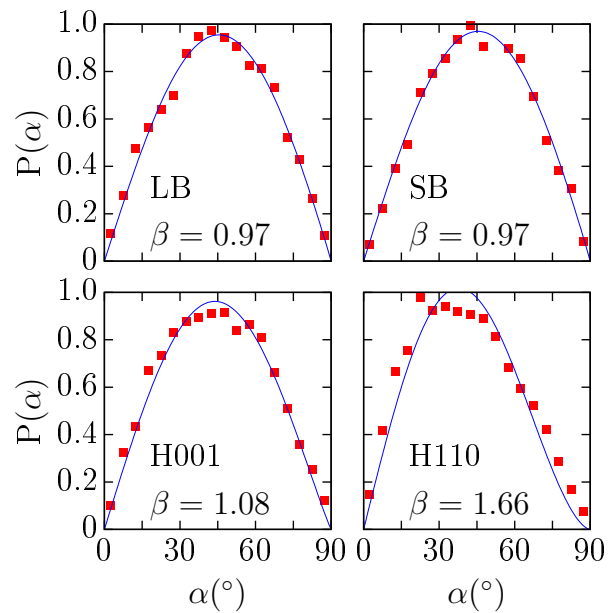} 
\caption{(Color online) Angular distribution of the molecules desorbed from the four adsorption wells for a laser fluence of $F=200$ J/m$^2$. Red squares show the results of the dynamics. The blue line is obtained by fitting the data to the following function: $P(\alpha)=(\beta+1)\sin \alpha \cos^{\beta} \alpha$.  \label{ang}}
\end{figure}

In the following, we analyze the characteristics of the molecules desorbed from the different wells. 
The corresponding angular distributions of the desorbed molecules are shown in Fig.~\ref{ang}. These distributions are rather symmetrical around a desorbing angle of $45^\circ$ relative to the surface normal for all the adsorption wells. Nevertheless, in the case of molecules desorbed from the H110 site a slight tendency to desorb into directions closer to the surface normal is observed. We fit the obtained angular distribution to the velocity integrated flux-weighted Maxwell-Boltzmann distribution~\cite{Zimmermann1995127,peter_pccp} that gives $P(\alpha)=(\beta+1)\sin \alpha \cos^{\beta} \alpha$. The parameter $\beta$ is a measure of the alignment of the desorption flux. For large values of $\beta$ the flux is aligned normal to the surface and the distribution is narrow, while $\beta=1$ corresponds to a cosine distribution. As seen in Fig.~\ref{ang},
$\beta$ is practically one for desorption from the H100, LB, and SB wells and it is somewhat larger than one for the H110 well. These (small) values contrast with the values $\beta\gtrsim3$ obtained for the associative desorption of H$_2$ from Ru(0001)~\cite{peter_pccp}. In that case, the deviation from the cosine distribution was explained by the presence of a late barrier towards desorption, causing a channeling effect and narrow angular distribution.  
However, in our case, the potential energy defining the molecule-surface interaction is monotonically increasing from the wells to the vacuum region (see Fig. 5 of Ref.~\cite{ivor1}), which results in $\beta\sim1$.

\begin{figure}[ht]
\includegraphics[width=\columnwidth]{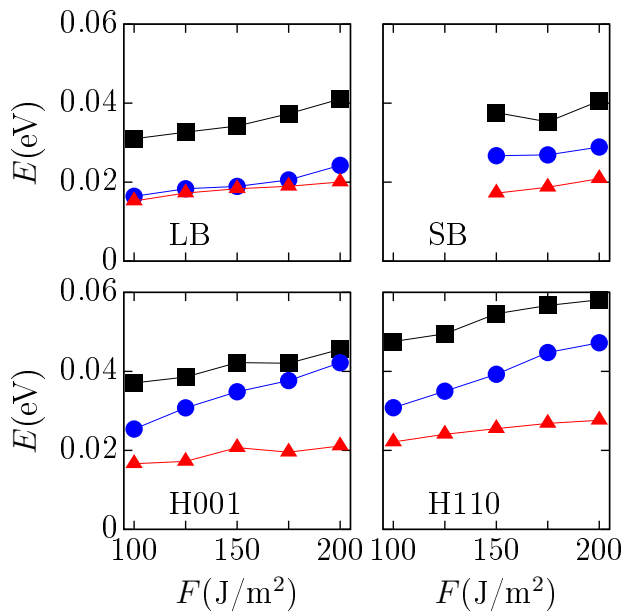} 
\caption{(Color online) Partition of the energy into translational (black squares), vibrational (blue circles), and rotational (red triangles) degrees of freedom of the molecules desorbed from the four wells as a the function of laser fluence $F$. Values for low fluences and SB well are not shown due to the poor statistics. \label{eqvi}}
\end{figure}

Next, we analyze how the energy of the desorbing molecules are partitioned in translational and internal (vibrational and rotational) degrees of freedom. The dependence of the translational, vibrational, and rotational energies of the molecules desorbed from the four adsorption wells as a function of the laser fluence is shown in Fig.~\ref{eqvi}. 
Equipartition of energy between the different degrees of freedom of a free diatomic molecule means that the values of its translational, vibrational, and rotational energies are ordered according to the ratio $3:2:2$~\cite{oxtoby2015principles}.
Figure~\ref{eqvi} shows that only in the case of LB, where desorption is dominantly phonon-mediated, the ideal thermal desorption is approximately fulfilled.
Deviations from the ideal ratio are already observed for desorption from the SB well, which is also a phonon-dominated process, and more clearly for H001 and H110 (both electron dominated).
In these three cases, and for all the laser fluences, the translational energy is the largest and the rotational energy is the lowest. Within a good approximation, a linear increase with the laser fluence of the vibrational and translational energies of the desorbed molecules is observed. This is considered to be one of the hallmarks of DIMET~\cite{peter_rew3}.

Finally, it is worth noting that we also observe few dissociation events for molecules initially adsorbed in the H110 well.  
This is a very unlikely process that has only been observed at the highest fluences ($F>$ 175 J/m$^2$) and with probabilities lower than $10^{-4}$. In these conditions it is not possible to perform a more detailed analysis of the process. Still, the occurrence of dissociation events is an interesting result considering that the energy barrier to dissociation is $0.57$ eV~\cite{ivor1}, significantly larger than the well depth of $-0.21$~eV.

%
%
\section {Summary}\label{summary}
In summary, we have extended the approach of Ref.~\cite{peter_pccp} to
simulate the multidimensional dynamics of a molecule adsorbed on a metal
surface excited by an ultrashort laser pulse by including surface
movement (phonons) via the GLO model. This allows us to treat
simultaneously the laser induced electron and phonon excitations and their effect on the dynamics of the eventually desorbing molecule.

Using this new approach we have studied the laser induced desorption of
O$_2$ from Ag(110). An interesting feature of this system is that it
possesses four distinct molecular adsorption wells. This enables us to study how the
desorption mechanisms are connected to the properties of the adsorption
configuration. In general, we find that the effect of the laser-heated
phonons in this system cannot be
disregarded. Importantly, the phonon contribution to the desorption
yield can be either positive or negative depending on the
adsorption site. More precisely, when the molecule is initially adsorbed
on the bridge sites inclusion of phonons increases the desorption probability.
In fact, for these sites, coupling of the molecule to the phonon
excitations constitutes the main desorption mechanism. However, for
molecules adsorbed on the hollow sites not only the electronic channel
is the dominant mechanism, but inclusion of phonons reduces the
desorption probabilities because they take energy from the excited
molecule. The subsequent reductions of the desorption yields can be
rather high, in the range of a factor $2-7$, depending on the laser
fluence. These observations are rationalized in terms of the distances
from the surface at which the adsorption sites are located and the
subsequent values of the electronic density in their surroundings. Hollow sites are
closer to the surface than bridge sites and, consequently, in regions of
higher electronic density. For this reason the electron channel
dominates desorption in the former and the phonon channel in the latter.

Our results also suggest which desorption mechanism will be dominant in systems 
that present both physisorbed and chemisorbed species. Since physisorbed molecules are located
in low electronic density regions their desorption behavior is expected to be similar to the one we obtain for the bridge sites, whereas for chemisorbed states our findings for hollow sites apply.

\section{Acknowledgement}
Fruitful discussions with G. F\"{u}chsel (Leiden) are gratefully acknowledged. I.L., M.A. and J.I.J. acknowledge the Basque Departamento de Educaci\'{o}n, Universidades e Investigaci\'{o}n, the University of the Basque Country UPV/EHU (Grant No. IT-756-13) and the Spanish Ministerio de Econom\'{i}a y Competitividad (Grant No. FIS2013-48286-C2-2-P). P.S. acknowledges support by the Deutsche Forschungsgemeinschaft (DFG) through project Sa 547/8. Computational resources were provided by the DIPC computing center.


%


\end{document}